\documentstyle[aps,twocolumn]{revtex}

\begin{document}
\title{Implementation of a quantum algorithm to solve Bernstein-Vazirani's parity
problem without entanglement on an ensemble quantum computer}
\author{{\small Jiangfeng Du\thanks{%
Email: djf@ustc.edu.cn}, Mingjun Shi, Xianyi Zhou, Yangmei Fan, BangJiao Ye,
Rongdian Han}}
\address{Laboratory of Quantum Communication and Quantum Computation,\\
Department of Modern Physics,\\
University of Science and Technology of China, Hefei, 230026, P.R.China}
\author{{\small \ Jihui Wu}}
\address{Laboratory of Structure Biology, University of Science and Technology of\\
China, Hefei, 230026, P.R.China}
\date{Nov. 11, 2000}
\maketitle

\begin{abstract}
Bernstein and Varizani have given the first quantum algorithm to solve
parity problem in which a strong violation of the classical imformation
theoritic bound comes about. In this paper, we refine this algorithm with
fewer resource and implement a two qubits algorithm in a single query on an
ensemble quantum computer for the first time.

\end{abstract}

\medskip

Associated with the model of quantum computer\cite{1,2}, a variety of
quantum algorithm has been proposed\cite{3,4,5,6,7}. In the theoretical
view, these algorithms have relevance to the entanglement phenomena, the
peculiar quantum property first identified by Erwin Schroedinger\cite{8},
which is invoked as the mechanism for the speedup of quantum computing over
their classical counterpart\cite{9}. Until recently these algorithms were
only of theoretical interest, as it proved extremely difficult to build a
quantum computer. In the last few years, however, there has been substantial
progress based on nuclear magnetic resonance (NMR)\cite{10}. Up to now some
simple quantum algorithms have been realized step by step on NMR quantum
computer, including Deutsch's algorithm\cite{11,12,13,14,15,16,17}, Grover's
algorithm\cite{18,19,20,21,22} and ordering find's algorithm\cite{23}.
However, a sharp criticism has been proposed by Braunstein et al. that NMR
experiments have not actually realized quantum algorithm because at each
time step the state of the system can be described as a probabilistic
ensemble of unentangled quantum states\cite{24}. On the other hand, some
scientists believe that for a specific quantum algorithm the power of
quantum computer derives from quantum superposition and parallelism, other
than entanglement\cite{25,26,27,28}.

The problem we considered in this paper is the parity problem about a
database $A$ that contains an arbitrary $n$-bit string $a$. The answer to
queries represented by $n$-bit string $x$ to the database is the parity of
the bits common to $x$ and $a$ given by $\left( a,x\right) =a\cdot x.$ Note
that the problem is to determine $a$ in its entirety, not to merely
determine the parity of $a$. the classical determination of $n$-bit string $%
a $ requires at least $n$ query operations (since n-bit string $a$ contains $%
n$ bits of information and each classical evaluation of query operation
yields a single bit of information). Bernstein and Vazirani have given the
first quantum algorithm in which $n$-bit string $a$ can be determined in
only two queries to the database\cite{6}. But by preparing the output one
bit register in an initial superposition $\frac 1{\sqrt{2}}\left( \left|
0\right\rangle -\left| 1\right\rangle \right) $, the algorithm can be
simplified to comprise a single query\cite{29}. Terhal and Smolin
rediscovered this algorithm, which was underappreciated, to solve binary
problems and coin-weighing problems effectively\cite{30}.

In this paper, based on Bersntein-Vazirani's parity problem\cite{6}, we
proposed a scheme to solve this problem by using less physics resource than
the previous algorithm\cite{28,29}, but without loss of effectivity.
Further, we demonstrated this algorithm on a two-qubit NMR quantum computer
for the first time.

\medskip

Bernstein and Vazirani's parity problem can be described as a function $%
f:\left\{ 0,1\right\} ^n\rightarrow \left\{ 0,1\right\} $, which is of the
form $f_a\left( x\right) =a\cdot x\equiv \left( \sum_{i=1}^na_ix_i\right) 
\mathop{\rm mod}%
2$, where n-bit strings $a,x\in \left\{ 0,1\right\} ^n$, $a_i$ and $x_i$ are
the $i^{th}$ bits of $a$ and $x$, and $a\cdot x$ denotes the bitwise AND (or
mod $2$ scalar product) $a\cdot x\equiv \left( a_1\Lambda x_1\right) \otimes
\left( a_2\Lambda x_2\right) \otimes \cdots \otimes \left( a_n\Lambda
x_n\right) $, the answer is to find $n$-bit string $a$. The previous quantum
algorithm\cite{28,29} to solve this problem theoretically by a pair of
registers $\left( x,b\right) $ , where $x\in \left\{ 0,1\right\} ^n,$ $b\in
\left\{ 0,1\right\} $. The quantum network to implement the algorithm is
shown in figure 1, the $n+1$ qubits register $\left( x,b\right) $ start in
the state $\left| x\right\rangle \left| b\right\rangle =\left( \left|
0\right\rangle \right) ^n\frac 1{\sqrt{2}}\left( \left| 0\right\rangle
-\left| 1\right\rangle \right) $. The function $f_a\left( x\right) =a\cdot x$
is designed within a unitary operator $U_f$ which denotes the transform

\begin{equation}
\left| x\right\rangle \left| b\right\rangle \stackrel{U_f}{\rightarrow }%
\left| x\right\rangle \left| y\oplus f\left( x\right) \right\rangle \equiv
\left| x\right\rangle \left| b\oplus \left( a\cdot x%
\mathop{\rm mod}%
2\right) \right\rangle
\end{equation}
The Hadmard Gate $H$ denote the transform

\begin{equation}
\begin{array}{l}
\left| 0\right\rangle \rightarrow \frac 1{\sqrt{2}}\left( \left|
0\right\rangle +\left| 1\right\rangle \right) \\ 
\left| 1\right\rangle \rightarrow \frac 1{\sqrt{2}}\left( \left|
0\right\rangle -\left| 1\right\rangle \right)
\end{array}
,~H=\frac 1{\sqrt{2}}\left( 
\begin{array}{cc}
1 & 1 \\ 
1 & -1
\end{array}
\right)
\end{equation}
If we apply $n$ Hadmard gates $H^{\left( n\right) }=H\otimes H\otimes \cdots
\otimes H$ in parallel to $n$-qubit, then the $n$-qubit state transforms as $%
H^{\left( n\right) }:\left| x\right\rangle \rightarrow \frac 1{\sqrt{2^n}}%
\sum\limits_{y=0}^{2^n-1}\left( -1\right) ^{x\cdot y}\left| y\right\rangle $%
. Therefore, acting on the input $\left( \left| 0\right\rangle \right) ^n%
\frac 1{\sqrt{2}}\left( \left| 0\right\rangle -\left| 1\right\rangle \right) 
$, with $f_a\left( x\right) =a\cdot x$ and $\frac 1{2^n}\sum%
\limits_{y=0}^{2^n-1}\left( -1\right) ^{a\cdot x+x\cdot y}=\delta _{ay}$, we
can evaluate the state of the input register as

\begin{equation}
\begin{array}{l}
\left( \left| 0\right\rangle \right) ^n\left| 1\right\rangle \stackrel{%
H^{\left( n+1\right) }}{\rightarrow }\frac 1{\sqrt{2^n}}\sum%
\limits_{x=0}^{2^n-1}\left| x\right\rangle \otimes \frac{\left|
0\right\rangle -\left| 1\right\rangle }{\sqrt{2}} \\ 
\stackrel{U_f}{\rightarrow }\frac 1{\sqrt{2^n}}\sum\limits_{x=0}^{2^n-1}%
\left( -1\right) ^{f_a\left( x\right) }\left| x\right\rangle \otimes \frac{%
\left| 0\right\rangle -\left| 1\right\rangle }{\sqrt{2}} \\ 
\stackrel{H^{\left( n\right) }}{\rightarrow }\frac 1{\sqrt{2^n}}%
\sum\limits_{x=0}^{2^n-1}\sum\limits_{y=0}^{2^n-1}\left( -1\right) ^{a\cdot
x+x\cdot y}\left| y\right\rangle \otimes \frac{\left| 0\right\rangle -\left|
1\right\rangle }{\sqrt{2}} \\ 
\equiv \left| a\right\rangle \otimes \frac{\left| 0\right\rangle -\left|
1\right\rangle }{\sqrt{2}}
\end{array}
\end{equation}
It is obviously that one could execute the quantum network once and measure
the $n$-qubit input register, finding the $n$-bit string $a$ in function $%
f_a\left( x\right) =a\cdot x\equiv \left( \sum_{i=1}^na_ix_i\right) 
\mathop{\rm mod}%
2$ with probability one.

Our refined version of Bernstein-Vazirani algorithm uses $n$ qubits, rather
than $n+1$ qubits, to find $n$ bits string $a$. A quantum circuit for this
refined quantum algorithm is shown in figure 2. To compare with the original
algorithm shown in figure 1, the one-qubit work register $b$ is removed
because it is redundant in the sense that its state does not change. To do
so, the binary function $f_a\left( x\right) =a\cdot x$ encoded in a $n+1$
qubits unitary transformation $U_f$ was changed into the $n$ qubits
propagator $U_a$ such that

\begin{equation}
\left| x\right\rangle \stackrel{U_a}{\rightarrow }\left( -1\right)
^{f_a\left( x\right) }\left| x\right\rangle
\end{equation}
, and the unitary transformation $U_a$ can be decomposed as direct products
of single qubit operators

\begin{equation}
U_a=U^1\otimes U^2\otimes \cdots \otimes U^i\otimes \cdots \otimes
U^{n-1}\otimes U^n
\end{equation}

\begin{equation}
U^i=\{ 
\begin{array}{l}
I,\quad \\ 
\sigma _z,\quad
\end{array}
\begin{array}{l}
a_i=0 \\ 
a_i=1
\end{array}
,\quad I=\left( 
\begin{array}{ll}
1 & 0 \\ 
0 & 1
\end{array}
\right) ,\sigma _z=\left( 
\begin{array}{ll}
1 & 0 \\ 
0 & -1
\end{array}
\right)
\end{equation}

The first Hadmard Gate $H^{\left( n\right) }=H\otimes H\otimes \cdots
\otimes H$ takes $\left| \psi _0\right\rangle =\left( \left| 0\right\rangle
\right) ^n$ to $\left| \psi _1\right\rangle =\frac 1{\sqrt{2^n}}%
\sum\limits_{x=0}^{2^n-1}\left| x\right\rangle $. After the unitary
transformation $U_a$ responds to this quantum query, the state is $\left|
\psi _2\right\rangle =$ $\frac 1{\sqrt{2^n}}\sum\limits_{x=0}^{2^n-1}\left(
-1\right) ^{f_a\left( x\right) }\left| x\right\rangle =$ $\frac 1{\sqrt{2^n}}%
\sum\limits_{x=0}^{2^n-1}\left( -1\right) ^{a\cdot x}\left| x\right\rangle $%
. The final Hadmard Gate $H^{\left( n\right) }$ outputs $\left| \psi
_3\right\rangle =$ $\frac 1{\sqrt{2^n}}\sum\limits_{x=0}^{2^n-1}\sum%
\limits_{y=0}^{2^n-1}\left( -1\right) ^{a\cdot x+x\cdot y}\left|
y\right\rangle =\left| a\right\rangle $. Whereupon measuring the whole $n$
qubits register identifies $a$ with probability 1 (the output states for
different $a$'s are orthogonal).

Quantum entanglement and quantum interfere are usually thought to be the key
gradient in quantum algorithm and the reason why quantum algorithm exceed
classical algorithm. But in the above refined quantum algorithm, there is no
entanglement in it. The initial state is $\left| \psi _0\right\rangle
=\left( \left| 0\right\rangle \right) ^n$, which is obvious separable. After
the Hadamard transformation, the state is $\left| \psi _1\right\rangle =%
\frac 1{\sqrt{2^n}}\left( \left| 0\right\rangle +\left| 1\right\rangle
\right) \otimes \left( \left| 0\right\rangle +\left| 1\right\rangle \right)
\otimes \cdots \otimes \left( \left| 0\right\rangle +\left| 1\right\rangle
\right) $. Performed by query operations $U_a$, the state becomes $\left|
\psi _2\right\rangle =\frac 1{\sqrt{2^n}}\left( \left| 0\right\rangle
+e^{i\pi a_0}\left| 1\right\rangle \right) \otimes \left( \left|
0\right\rangle +e^{i\pi a_1}\left| 1\right\rangle \right) \otimes \cdots
\otimes \left( \left| 0\right\rangle +e^{i\pi a_n}\left| 1\right\rangle
\right) $.And the states after the second Hadamard transformation is the
output state $\left| \psi _3\right\rangle =\left| a_0\right\rangle \left|
a_1\right\rangle \cdots \left| a_n\right\rangle $. In the whole procedure,
the state is tensor products of the states of the individual qubits, so it
is unentanlged. And because the operators in the algorithm ($H^{\left(
n\right) }$ , $U_a$ and $H^{\left( n\right) }$) are also tensor product of
the individual local operators on these qubits; $H^{\left( n\right)
}=H\otimes H\otimes \cdots \otimes H$ and $U_a=U^1\otimes U^2\otimes \cdots
\otimes U^i\otimes \cdots \otimes U^{n-1}\otimes U^n$. Such a unitary
transformation cannot change the entanglement of a state.

\medskip

Experimentally, this quantum algorithm without entanglement was implemented
using the nuclear spins of the two hydrogen atoms in a deuterated cytosine
molecule. $\left| 0\right\rangle \left( \left| 1\right\rangle \right) $
describes the spin state aligned with (against) an externally applied,
strong static magnetic field $B_0$ in the $\widehat{z}$ direction. the
reduced Hamiltonian for this two-spin system is to an excellent
approximation given by

\begin{equation}
H=\omega _AI_z^A+\omega _BI_z^B+2\pi JI_z^AI_z^B
\end{equation}
where the first two terms describe the free precession of spin $A$ and $B$
of two hydrogen atoms about $B_0$ with frequencies $\frac{\omega _A}{2\pi }%
\approx $ $\frac{\omega _B}{2\pi }\approx 500Mhz$, and the chemical shift $%
\left| \frac{\omega _A}{2\pi }-\frac{\omega _B}{2\pi }\right| =765Hz$ enable
us to address each spin (act as qubit) individually. $I_z^A$ is the angular
momentum operator in the $+\widehat{z}$ direction for $A$, The third term
describes a scalar spin-spin coupling of the two spins of $J\approx 7.17Hz.$
As we know, pulsed radiofrequency (RF) electromagnetic fields, oriented in
the $\widehat{x}-\widehat{y}$ plane perpendicular to static magnetic field $%
B_0$, selectively address either $A$ or $B$ by oscillating at frequency $%
\omega _A$ and $\omega _B$. For example, a RF pulse along $\widehat{y}$
rotates a spin about that axis by an angle $\theta $ proportional to $\theta
\approx tP$, the product of the pulse duration $t$ and pulse power $P$. In
this paper, we shall let $R_y^A\left( \theta \right) $ denote $\theta $
rotations act on spin $A$ about $\widehat{y}$, $R_x^{AB}\left( \theta
\right) $ denote $\theta $ rotations act on spin $A$ and $B$ about $\widehat{%
x}$ simultaneously, and so forth; superscripts will identify which spin the
operation acts upon, subscripts denote which axis an RF pulse rotates a spin
about.

Experiments are conducted at room temperature and pressure on Bruker Avance
DMX-500 spectrometer in Laboratory of Structure Biology, University of
Science and Technology of China. A quantum circuit for implementing this
algorithm on a two qubit NMR quantum computer is shown in figure 2 with $n=2$%
. In our experiment, pairs of Hadmard gates were replaced by an NMR
pseudo-Hadmard gate $h$ (a $90_y^o$ rotation) and its inverse $h^{-1}$\cite
{31}. An input peseudopure state $\psi _0=\left| 00\right\rangle $ was
generated using the approach of Cory $et$ $al.$\cite{32,33}. This is
implemented as $R_x^B\left( \pi /3\right) -G_z-R_x^A\left( \pi /4\right)
-\tau -R_y^A\left( -\pi /3\right) -G_z$, to be read from left to right,
where $G_z$ is the pulsed field gradient along the $\widehat{z}$ axis to
annihilate all transverse magnetizations, dashes are for readability only,
and $\tau $ represents a time interval of $1/\left( 2J\right) \approx
69.735ms$.

The pair of pseudo-Hadmard gates $h$ and $h^{-1}$ could be easily
implemented by two hard pulses denoted as $R_{-y}^{AB}\left( \pi /2\right) $
and $R_y^{AB}\left( \pi /2\right) $, the typical pulse lengths were $10-20us$%
. All unitary transformation $U_a$ corresponding to the query of four
possible 2-bit string$\ a=\left\{ 00,01,10,11\right\} $ in function $%
f_a\left( x\right) =a\cdot x$ could be denoted as $U_{00}=I^A\otimes
I^B,U_{01}=I^A\otimes \sigma _z^B,U_{10}=\sigma _z^A\otimes
I^B,U_{11}=\sigma _z^A\otimes \sigma _z^B$. The $U_{00}$ transformation
corresponds to the unity operation or ``do nothing''. $U_{01}$ and $U_{10}$
transform are separately achieved by applying $R_z^B\left( \pi \right) $
rotation selectively on the second qubit $B$ and $R_z^A\left( \pi \right) $
rotation on the first qubit $A.$ The soft z pulse was implemented by the
time evaluation under the Hamiltonian of Eq. (6) with refocusing $\pi $
pulses applied at suitable times during the evolution period. Since the
refocusing $\pi $ pulse has the effect of time reversal, it can be used to
make one term in the Hamiltonian evolve while the other terms ``freeze''\cite
{34,35}. In our experiment, we extended these method to realize the soft
pulses $R_z^A\left( \pi \right) $ and $R_z^B\left( \pi \right) $ separately
as 
\begin{eqnarray*}
\tau _1/4-R_x^B\left( \pi \right) -\tau _1/2-R_{-x}^B\left( \pi \right)
-\tau _1/4,\ \omega _A\tau _1 &=&\pi \\
\tau _2/4-R_x^A\left( \pi \right) -\tau _2/2-R_{-x}^B\left( \pi \right)
-\tau _2/4,\ \omega _B\tau _2 &=&\pi
\end{eqnarray*}
the axes of successive $\pi $ pulses were chosen in the way to cancel
imperfections of soft pulses. $U_{11}$ corresponds to a $\pi $-rotation $%
R_z^{AB}\left( \pi \right) $ about the axis $\widehat{z}$ of both qubits, up
to a global phase factor. Global phase changes are not detectable in NMR and
are hence ignored for the purpose of experiment. This non-selectively $%
\widehat{z}$-rotation was implemented using a composite-pulse sandwich, as a
set of $\widehat{x}$ and $\widehat{y}$ axes $R_{-y}^{AB}\left( \pi /2\right)
-R_x^{AB}\left( \pi \right) -R_y^{AB}\left( \pi /2\right) .$

The result of this algorithm are shown in figure 3. Five aspects are shown:
a reference spectrum acquired using a single pulse $R_y^{}\left( \pi
/2\right) $, and spectra acquired from the same computer implementing the
query algorithm for each of the four possible functions, $f_a$. Each
spectrum consists of two closely spaced pair of lines: each pair of lines
corresponds to a single qubit, while the barely visible splitting within
each pair arises from the spin-spin coupling, $J$. To improve the appearance
of the spectra the final detection pulse was proceeded by a magnetic field
gradient pulse, which acts to dephase the majority of any error terms which
might occur. The reference spectrum (a) corresponds to the computer being in
state $\left| 00\right\rangle $, and the phase of this spectrum was adjusted
so that both lines are in positive absorption phase(that is, pointing
upwards). The same phase correction was then applied to the other four
spectra, allowing positive absorption lines to be interpreted as qubits in
state $\left| 0\right\rangle $, while negative absorption lines can be
interpreted as qubits in $\left| 1\right\rangle $. The left hand pair of
lines arises from the first spin, and thus corresponds to the first qubit,
while the right hand pair of lines corresponds to the second qubits. Thus,
spectrum (b)-(e) correspond to $\left| a\right\rangle =\left|
00\right\rangle $, $\left| a\right\rangle =\left| 01\right\rangle $, $\left|
a\right\rangle =\left| 10\right\rangle $ and $\left| a\right\rangle =\left|
11\right\rangle $ respectively$.$ It is clear that our implementation of a
two bit parity problem using a single query leaves the computer in a final
state, much as expected.

\medskip

To summarize, we have presented a refined version of Bernstein and
Vazirani's quantum algorithm to solve the parity problem with fewer resource
and implement this algorithm in a single query on an ensemble quantum
computer for the first time. This algorithm, which gives a strong violation
of the classical information theoretic bound\cite{30} and a clear separation
between the quantum and classical difficulty of the problem\cite{6}, reduces
the number or queries all the way from $n$ to $1$. It is obviously that in
this algorithm there is no either entangle state or entangle transformation
but only the concept of coherent superposition is exploited, to prepare ``in
parallel'' an input state which is a superposition of all possible classical
inputs. Our algrithm and its experimental realization demenstrate that the
superposition principle brings about more effective and concise procedure
even if the entanglement phenomena do not occur. As we all know, some
quantum algrithms have relevence with entanglement\cite{3,7}, but some other
not\cite{25,26,27,28}, so it is meaningful to know the role of entanglement
in quantum algrithms, i.e., the relationship between the entanglement and
the complexity of algrithm.

\medskip

{\bf ACKNOWLEDGMENTS}

This project is supported by the National Nature Science Foundation of China
(10075041 and 10075044) and the Science Foundation of USTC for Young
Scientists.

\medskip

\bigskip

{\bf Figure caption:}

figure 1: A (schematic) quantum circuit implementing Bersntein-Vazirani's
parity problem in single query. The upper $n$ line corresponds to $n$ qubits
of register $X$, while the lower one line corresponds to one qubit of
register $b$.

figure 2: The (schematic) refined version of the quantum circuit shown in
figure 1, note that one qubit register $b$ is removed by changing the
unitary operator.

figure 3: Experimental spectra from our two-qubit NMR quantum computer.
spectrum (a) corresponds to the initial state $\left| 00\right\rangle $ of
register $x$, and act as a reference spectrum; while spectra (b)-(e) were
acquired from the same computer implementing the refined algorithm to solve
parity problem in a single query, and determine each of the four possible $2$%
-bit string $a$: (b) $a=00$, (c) $a=01$, (d) $a=10$, (e) $a=11$.

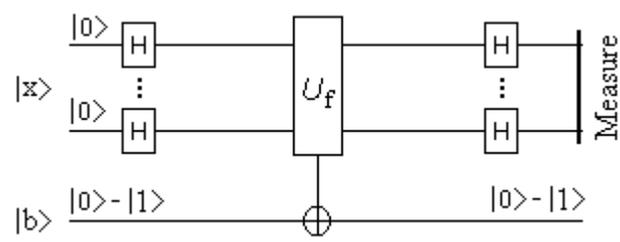
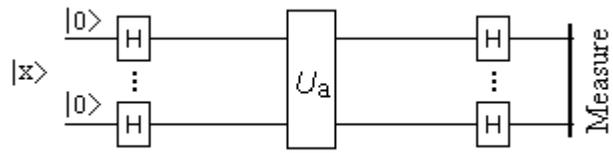
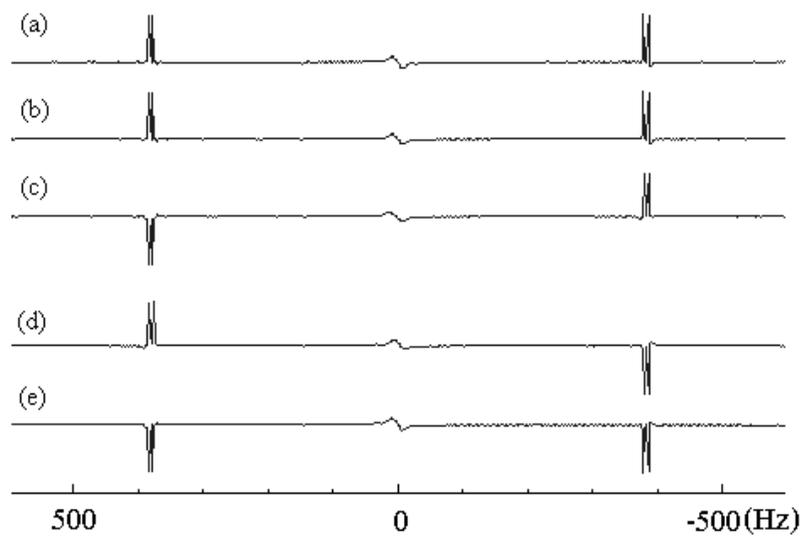
\end{document}